\documentclass[12pt,letterpaper]{article}
\usepackage[left = 1 in, right = 1 in, top = 1 in, bottom = 1 in]{geometry}
\usepackage{amsmath}
\usepackage{amssymb} 
\usepackage{graphicx}
\usepackage{setspace}
\usepackage{float}        
\usepackage{siunitx}
\usepackage{xspace}
\usepackage{xcolor}
\usepackage{soul}
\usepackage{listings} 

\usepackage{hyperref}
\usepackage[style=nature,hyperref=true]{biblatex}
\newcommand{\varnamesymbol}[3]{\ensuremath{#1_{\text{#2}}^{\text{#3}}}\xspace}
\newcommand{\reffig}[1]{Fig.~\ref{#1}}

\newcommand{\e}[1]{\ensuremath{\times 10^{#1}}}
\newcommand{\Ohm}[0]{$\Omega$\xspace}
\newcommand{\micron}[0]{\textmu m\xspace}

\newcommand{\uA}[0]{\textmu A\xspace}

\newcommand{\us}[0]{\textmu s\xspace}

\newcommand{\Tc}[0]{\textit{T}\textsubscript{c}\xspace}
\newcommand{\Rs}[0]{\textit{R}\textsubscript{s}\xspace}

\newcommand{\trow}[0]{\varnamesymbol{t}{row}{}}
\newcommand{\tcol}[0]{\varnamesymbol{t}{col}{}}
\newcommand{\Lseries}[0]{\varnamesymbol{L}{s}{}}
\newcommand{\Lsnspd}[0]{\varnamesymbol{L}{snspd}{}}
\newcommand{\Ibias}[0]{\varnamesymbol{I}{bias}{}}
\newcommand{\Rbias}[0]{\varnamesymbol{R}{bias}{}}

\newcommand{\Rtc}[0]{\varnamesymbol{R}{tc}{}}
\newcommand{\Rshunt}[0]{\varnamesymbol{R}{sh}{}}
\newcommand{\taudead}[0]{\varnamesymbol{\tau}{dead}{}}

\begin{document}

\doublespacing
\pagenumbering{arabic} 


\title{A superconducting-nanowire single-photon camera with 400,000 pixels}

\author{
B. G. Oripov$^{1,2}$,
D. S. Rampini$^{1,2}$,
J. Allmaras$^3$, \\
M. D. Shaw$^3$,
S. W. Nam$^1$,
B. Korzh$^3$,
A. N. McCaughan$^1$
}

\date{
     \small
     $^1$National Institute of Standards and Technology, Boulder, Colorado 80305, USA\\%
     $^2$Department of Physics, University of Colorado, Boulder, Colorado 80309, USA\\%
     $^3$Jet Propulsion Laboratory, California Institute of Technology, 4800 Oak Grove Drive, Pasadena, California 91109, USA
 }
\maketitle


For the last 50 years, superconducting detectors have offered exceptional sensitivity and speed for detecting faint electromagnetic signals in a wide range of applications. These detectors operate at very low temperatures and generate a minimum of excess noise, making them ideal for  testing the non-local nature of reality~\cite{Giustina2015,belltest}, investigating dark matter~\cite{chilesdcr, Dixit2021},  mapping the early universe~\cite{Riechers2013,DeBernardis2000,Nones2012}, and performing quantum computation~\cite{Arrazola2021,Todaro2021,Madsen2022} and communication~\cite{Boaron2018, Wang2022,Grunenfelder2023, Li2023}. Despite their appealing properties, however, there are currently no large-scale superconducting cameras -- even the largest demonstrations have never exceeded 20 thousand pixels~\cite{Walter2020}. This is especially true for one of the most promising detector technologies, the superconducting nanowire single-photon detector (SNSPD)~\cite{Goltsman2001, Hadfield2009, Marsili2013}. These detectors have been demonstrated with system detection efficiencies of $98.0\%$~\cite{Dileep98}, sub-3-ps timing jitter~\cite{Boris3ps}, sensitivity from the ultraviolet (250\,nm)~\cite{Wollman2017} to the mid-infrared ($>$10\,\micron)~\cite{Verma2021}, and dark count rates below $6.2\e{-6}$ counts per second (cps)~\cite{chilesdcr}, but despite more than two decades of development they have never achieved an array size larger than a kilopixel~\cite{Wollman2019,tci1kpx}. Here, we report on the implementation and characterization of a 400,000 pixel SNSPD camera, a factor of 400 improvement over the previous state-of-the-art. The array spanned an area $4\times2.5$\,mm with a $5\times5$\,\micron resolution, reached unity quantum efficiency at wavelengths of \SI{370}{nm} and \SI{635}{nm}, counted at a rate of $1.1\times10^5$ cps, and had a dark count rate of $1.0\e{-4}$ cps per detector (corresponding to \SI{0.13}{cps} over the whole array). The imaging area contains no ancillary circuitry and the architecture is scalable well beyond the current demonstration, paving the way for large-format superconducting cameras with 100\% fill factors and near-unity detection efficiencies across a vast range of the electromagnetic spectrum.

Most superconducting sensors~\cite{Morozov2021}, such as microwave kinetic-inductance detectors (MKIDs) or hot-electron bolometers, produce continuously-valued outputs that lend themselves well to efficient readout schemes such as frequency multiplexing~\cite{Day2003a}. Superconducting nanowire single-photon detectors, however, are notoriously difficult to multiplex since they produce discrete pulses that are both low-amplitude and broadband.  These pulses are typically read out individually using one microwave readout cable per detector, unlike CCD-type sensors which accumulate charge at each pixel and can be serially interrogated with a single readout line. This per-pulse readout process be advantageous when photon-counting as it means there is no read noise, but severely limits the number of detectors that can be read out due to cryogenic thermal load limitations on the readout wiring.

To date, there have been many attempts to devise readout architectures that minimize the number of readout lines running from room temperature electronics to the detector chip. For example, one of the largest architectures to date used a time-of-flight measurement to measure photon position along the length of the detector and required only two readout lines~\cite{Zhao2017}. Unfortunately, since the entire imager was made from a single long nanowire, this architecture is susceptible to fabrication yield issues, as a single discontinuity can compromise the entire imager. Additionally, the large inductance of the long nanowire will inevitably lead to very long reset times and limit the maximum count rate. 

In row-column readout architectures, SNSPDs are arranged on an $N\times N$ grid~\cite{VarunRowCol}, enabling readout of $N^2$ pixels using $2N$ readout lines. The detection event is defined as correlated voltage pulses appearing on both row and column lines within a predefined time window. The row-column readout architecture was used to create the largest SNSPD array to date, a 1,024-pixel imager, but due to diminishing signal-to-noise (SNR) ratio with increasing pixel count, this approach cannot scale to much larger sizes~\cite{varun1kpx}. The thermal row-column scheme overcomes the SNR issue by keeping the row and column pixels electrically isolated and uses the local heat generated during the SNSPD detection process to generate correlated events on row and column channels~\cite{Jasonthermal}.

More recently, the thermally-coupled imager (TCI) architecture used a combination of thermal coupling to a superconducting bus and time-multiplexing along the readout bus, to achieve kilopixel imaging capability~\cite{tci1kpx}. This scheme used an asymmetric thermal coupling process to transmit pulses on a row or column detector to a readout bus, significantly reducing inter-detector crosstalk and enabling much greater scalability. The SNSPD array demonstrated in this work combines a thermal row-column sensor element with TCI readout in order to achieve its large degree of multiplexing. Additionally, the perpendicular orientation of the two nanowire layers enables polarization-insensitive optical cavity designs.  

\begin{figure}[h] 
    \centering
    \includegraphics[width=6.5in]{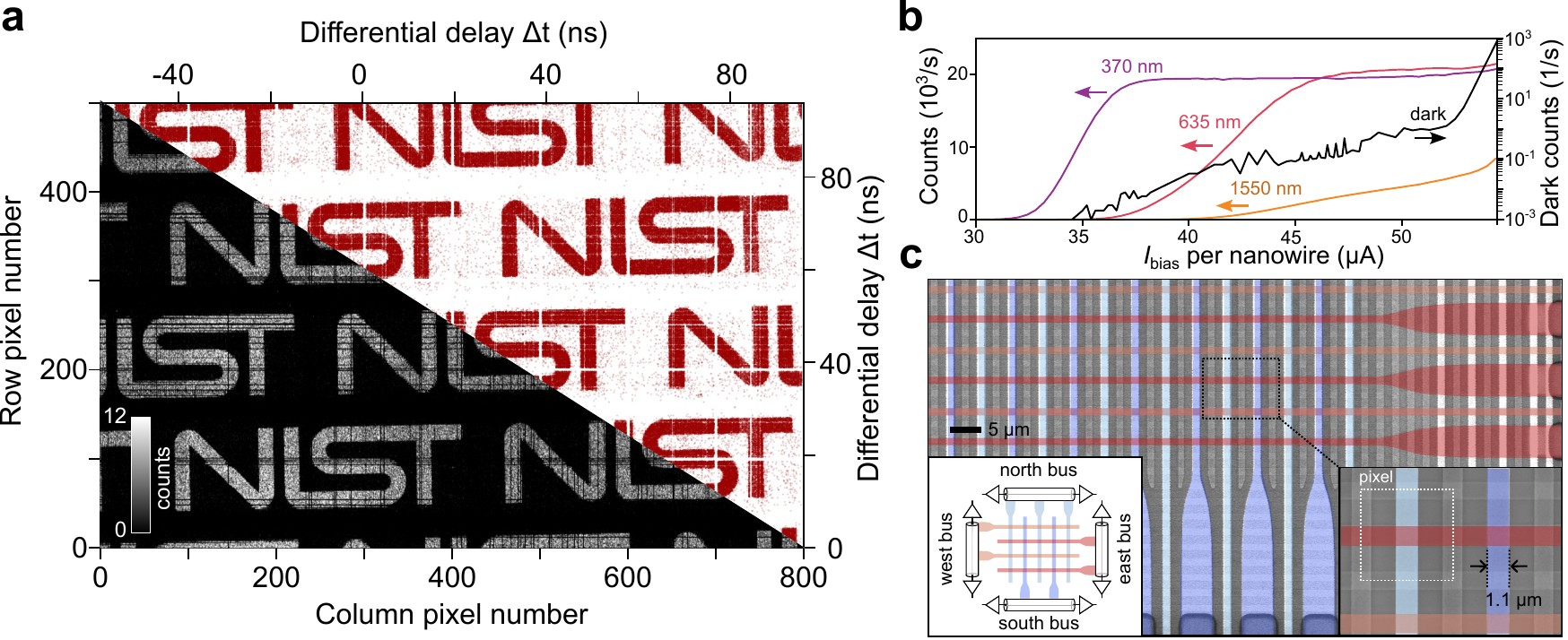}
    \caption{(a) Imaging with the $800\times500$ array at 370~nm.  Raw time-delay data from the buses are shown as individual dots in red, and binned 2D histogram data is shown in black and white.  (b) Count rate as a function of bias current for various wavelengths of light as well as dark counts. (c) False-color scanning electron micrograph of the lower-right corner of the array, highlighting the interleaved row and column detectors. (inset, lower left) Schematic diagram showing detector-to-bus connectivity.  (inset, lower right) Closeup showing 1.1~\micron detector width and effective $5\times5$~\micron pixel size.}
    \label{fig1-overview}
\end{figure}

The array, shown in~\reffig{fig1-overview}, was comprised of 800 row and 500 column detectors and was operated at a temperature of 0.8~K. The row and column detectors were made by patterning 1.1-\micron-wide wires from two independent 4-nm-thick WSi films with a \Tc of 3.4~K and kinetic inductance of \SI{250}{pH/sq}. Row and column detectors were spaced 5~\micron apart and arranged in a grid, giving a resolution of $5\times5$~\micron as shown in \reffig{fig1-overview}c.  There were four buses used for readout, requiring a total of 8 microwave coax lines, and the row and column detectors were interleaved between them as shown in~\reffig{fig1-overview}c. Each row or column detector was connected to a resistive thermal coupler, which contained a small heater that is thermally coupled to (but electrically isolated from) the readout bus by a thin, electrically insulating dielectric spacer.

The readout buses were made from the same layer of material as the column detectors and so have the same basic properties. The majority of the bus length had a width of 8~\micron that, in combination with the 50-nm-thick SiO$_2$ dielectric spacer and ground plane, formed a microstrip transmission line with approximately 50~\Ohm impedance. During operation, these sections of the bus contained a relatively low current density and were thus not directly photosensitive, allowing the readout bus to avoid generating false counts from stray photons. Below each heating element, the bus briefly narrowed to a 1.5-\micron-wide constriction with high current density, creating a heater-tron like device~\cite{Adamhtron}. The high current density ensured that every time the heater was activated during a detection event, it created a corresponding hotspot on the bus.

\begin{figure}[h] 
    \centering
    \includegraphics[width=6.5in]{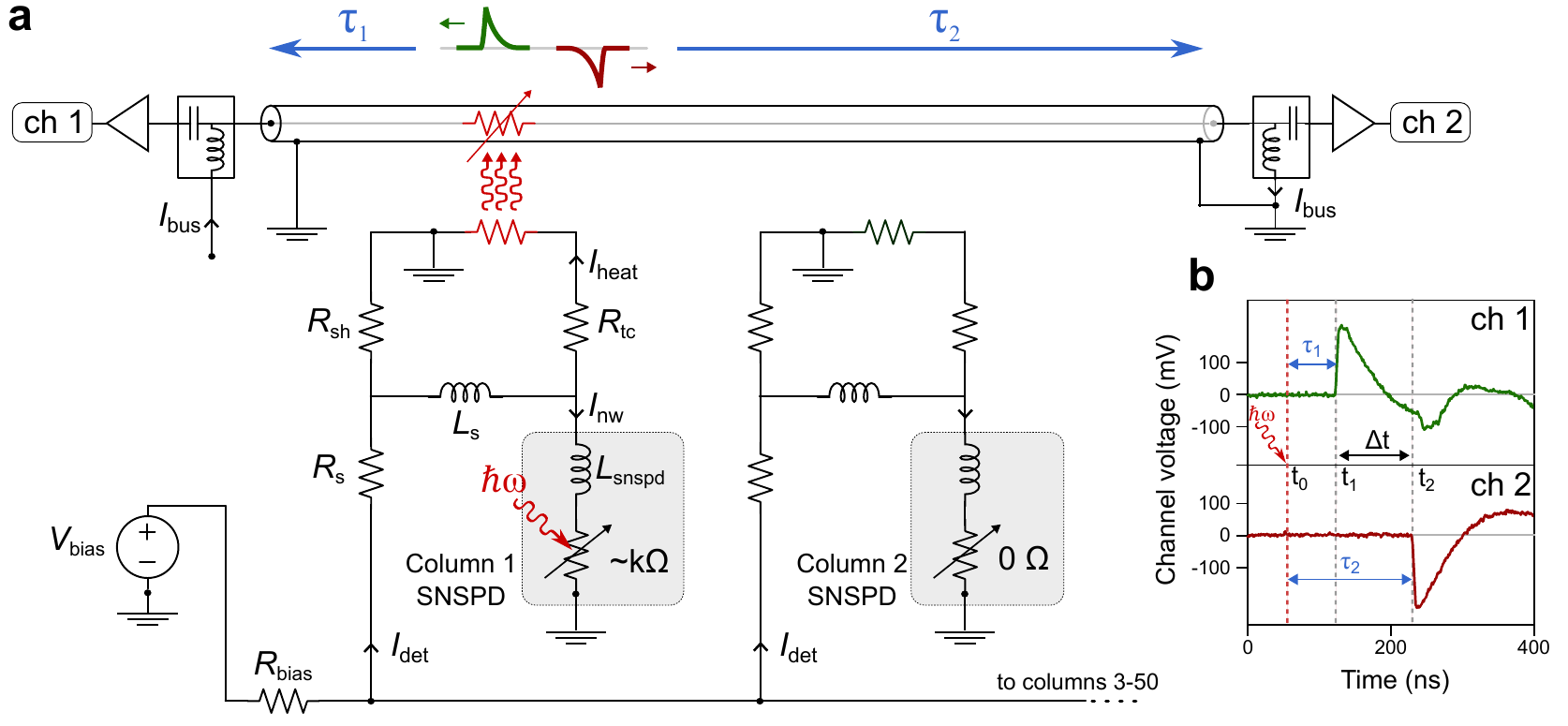}
    \caption{(a) Circuit diagram of a bus and one section of 50 detectors with ancillary readout components. SNSPDs are shown in the gray boxes, and all other components are placed outside of the imaging area. A photon that arrives at time $t_0$ has its location  determined by a time-of-flight readout process based on the time-of-arrival difference $t_2-t_1$. (b) Oscilloscope traces from a photon detection, showing the arrival of positive (green) and negative (red) pulses at times $t_1$ and $t_2$ respectively.}
    \label{fig2-circuit}
\end{figure}

Detectors were grouped into sections of 50 rows or 50 columns, and each section was biased independently. For each section, the bias-current was resistively distributed among the 50 detectors, sharing a common source and ground. \reffig{fig2-circuit} shows the circuit diagram for a single section comprising  50 column SNSPDs. When a photon was absorbed by a detector, a hot-spot formed on the detector with a peak value of a few k\Ohm. This resistance diverted the bias current out of the detector and into the heating element of the thermal coupler. The resulting phonons generated within the thermal coupler locally destroyed the superconducting state on the readout bus, creating two opposite-polarity voltage pulses that traveled down the readout bus as shown at the top and inset of \reffig{fig2-circuit}. Owing to the large kinetic inductance of the bus material, these microwave pulses propagated at 0.36\% of $c$ (1.10\e{6}~m/s), allowing relatively short lengths of bus (400~\micron) to  separate adjacent detectors while maintaining excellent distinguishability.

By measuring the time-difference between the arrival of these pulses at the ends of the buses, one can compute the location of the detection event and the corresponding row or column. As shown in \reffig{fig2-circuit}a, for a photon arriving at time $t_0$, the emitted positive pulse will arrive at the left readout at time $t_1 = t_0 + \tau_1$, and the negative pulse at $t_2 = t_0 + \tau_2$. Applying this process to both row and column readout, we were able to precisely identify the location where each photon was absorbed.  Given arrival times of $t_1$ and $t_2$ from a column bus and $t_3$ and $t_4$ from a row bus, we calculated the differential time-delays as $\Delta \tcol=t_1-t_2$, $\Delta \trow=t_3-t_4$. Similarly, the photon time of arrival can be calculated from $t_0=(t_1+t_2-\tau_b)/2$ where $\tau_b$ is the time-length of the bus. To ensure that these pulses were generated by the same single-photon detection event, we imposed the four-fold coincidence condition that the arrival time of all four pulses were within \SI{100}{ns} of each other.

To capture the image shown in \reffig{fig1-overview}a, we patterned a metallic mask on a glass substrate, placed it directly on top of the array, and flood illuminated the entire mask. We then recorded the pulse timings coming out of the readout buses with a timetagger, extracted the four-fold coincidence events, and computed $\Delta \tcol$ and $\Delta \trow$. No further post processing of the pulse data was performed and no 4-way coincidences were discarded. The raw differential-delay data shown in red corresponds closely to the black and white binned histogram image, excepting the regularly-spaced gaps in the differential-delay data that correspond to extra bus delay between adjacent detector sections. 

\reffig{fig1-overview}b shows photon count rate measurements for multiple wavelengths for one of the sections as a function of bias current. As evident from the figure, our detectors have large plateau regions for wavelengths \SI{370}{\nm} and \SI{635}{\nm}, indicating unity quantum efficiency over a wide range of bias current. At 370~nm, the wavelength for which this array was targeted, a bias current bias of 40~\uA yielded unity quantum efficiency and at this bias current we recorded 5 dark counts over a measurement period of 1,000~s. Since this measurement occurred over a section of 50 detectors, this corresponded to a dark count rate of $1\pm0.45\e{-4}$ cps per detector. In a setup that was better shielded optically, this value would likely be even lower, as the dark count curve in \reffig{fig1-overview}b suggests that the source of dark counts at this bias current were likely stray blackbody photons~\cite{dcrexpcurve}.

\begin{figure}[H] 
    \centering
    \includegraphics[width=70mm]{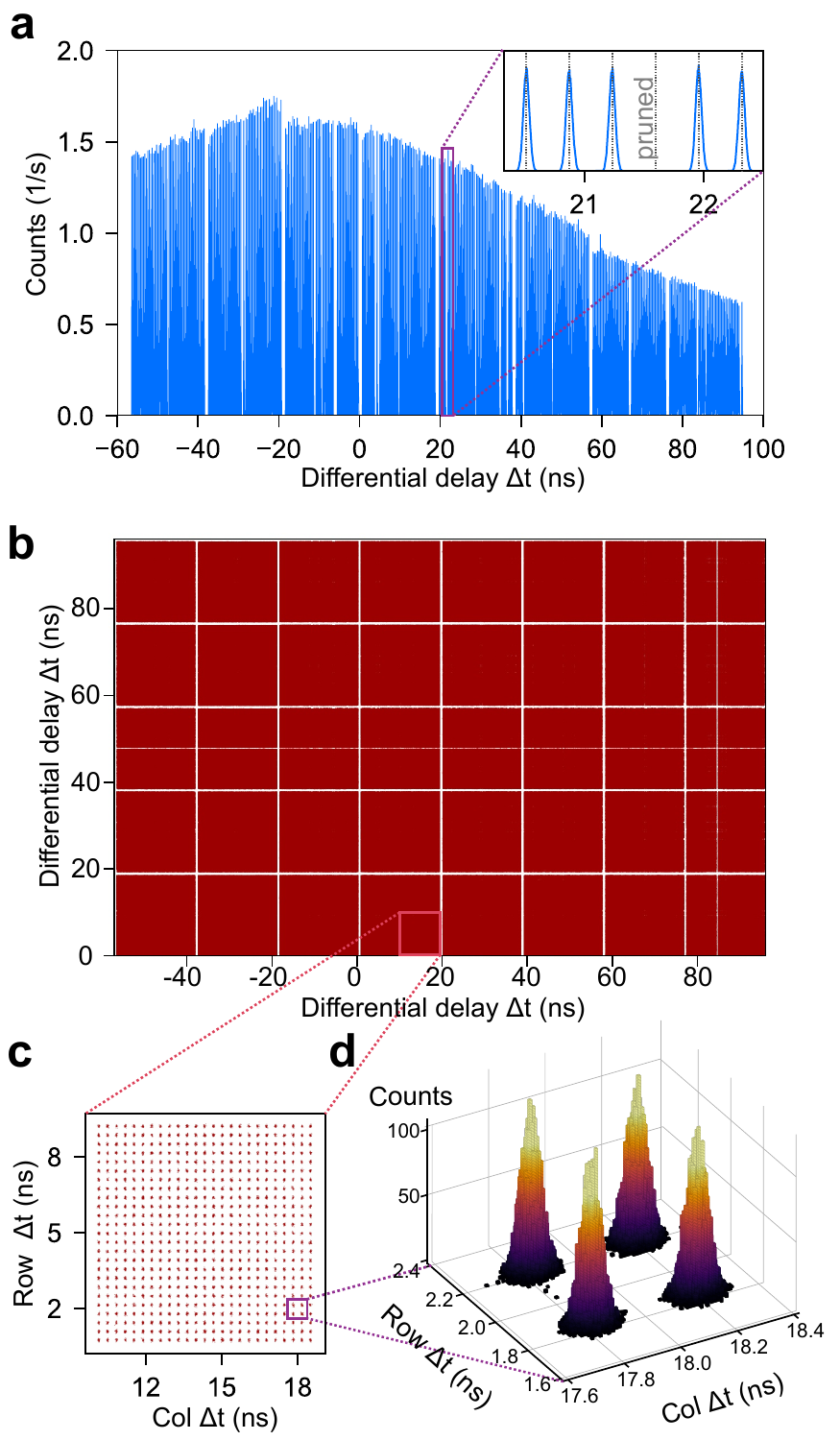}
    \caption{(a) Histogram of the pulse differential time-delays $\Delta t=t_1-t_2$ from the north bus during flood illumination with a Gaussian spot. All 400 detectors resolved clearly, with gaps indicating detectors that were pruned. (a, inset) Zoomed region showing that counts from adjacent detectors are easily resolvable, and no counts were generated by a pruned detector. (b) Plot of raw \trow and \tcol time-delays when flood illuminated at 370~nm. (c) Zoomed subsection of the array with $25 \times 25$ detectors. (d) Histogram of time-delays for a $2 \times 2$ detector subset with 10~ps bin size showing clear distinguishability between adjacent detectors.
    }
    \label{fig3-histograms}
\end{figure}

After an initial screening we observed that some detectors were defective, producing anomalously high dark count rates, overwhelming the bus, and obfuscating true counts. To resolve this, we applied a measure-and-prune process to eliminate the faulty detectors. We first determined exactly which detectors were anomalous by flood illuminating the array and collecting 1D histograms for each bus. We then identified any row or column detectors that produced excessive counts and disconnected them by etching through their bias wiring, effectively turning them off.  Within a single iteration of this pruning process, we were able to effectively remove all of the defective detectors (58 defective detectors out of 1300), resulting in the smooth Gaussian-spot shape of the histogram shown in \reffig{fig3-histograms}a. Identifying the offending detectors was straightforward -- the full-width half-max of the peaks was \SI{62}{ps} and the separation of the peaks had a median value of \SI{362}{ps}, giving a greater than $6$~FWHM distinguishability between neighboring rows or columns. We further examined the 2D uniformity of the array by flood illuminating it and examining the output of the buses. As shown in \reffig{fig3-histograms}b, the photon counts are tightly clustered in $\sim62$-ps-diameter circles, indicating that the readout process works to accurately position both the photon row and column simultaneously. 

One potential area for concern was that detected photons may trigger the detectors but not produce corresponding outputs on the buses, resulting in lost detection events. We characterized this photon-to-bus conversion efficiency with test structures, and found the minimum energy required to create a hotspot on the bus was 8.1\e{-17}~J. In the array, the detection process deposited $\frac{1}{2}\Lseries \Ibias$ = 1.2\e{-15}~J of energy on the heater, greatly exceeding the requirements of the thermal coupler. Correspondingly, during operation we observed large operating margins for the detectors and buses, indicating good uniformity and sensitivity of the thermal coupling process. When the detectors were biased between 30-55~\uA, any $\varnamesymbol{I}{bus}{}$ between 15-40~\uA produced a photon-to-bus conversion efficiency of approximately unity. We note that while the thermal coupling efficiency was high, an area for improvement will be the increasing the density of the layout for high fill-factors -- in the this work, only 13.7$\%$ of all photons detected resulted in a four-fold coincidence, due to the relatively large spacing between adjacent detectors.

\begin{figure}[H] 
    \centering
    \includegraphics[width=3.5in]{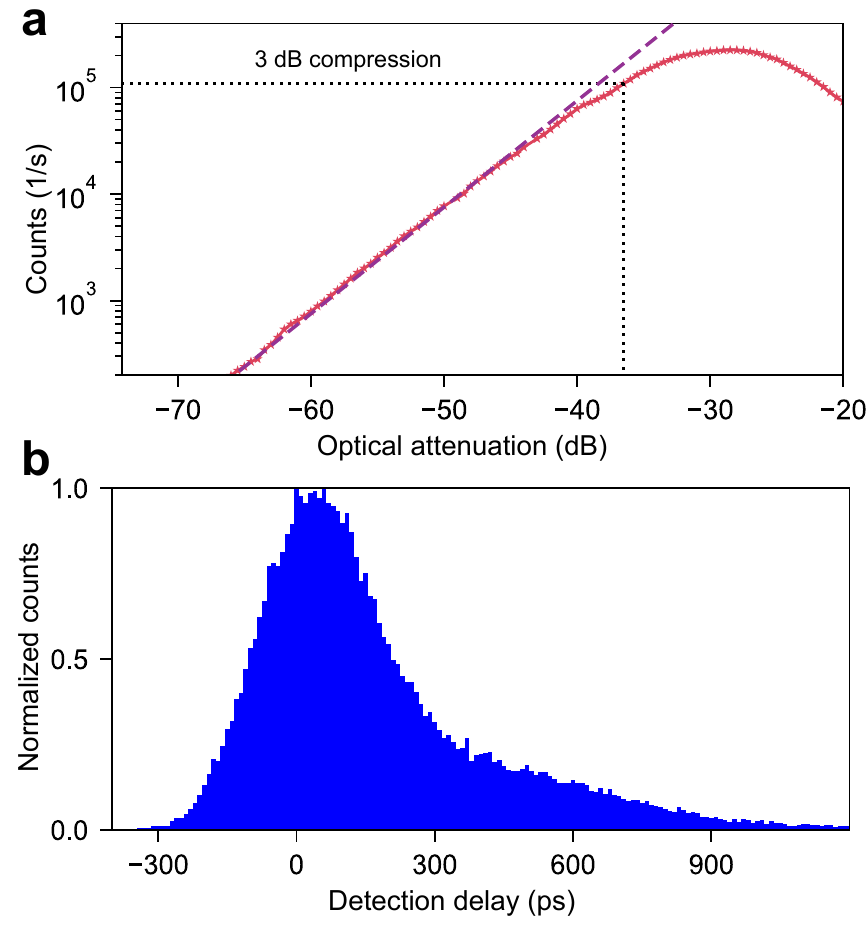}
    \caption{(a) Count rate versus optical attenuation for a section of detectors biased at 45~\uA per detector. The dashed purple line shows a slope of 1, with deviations from that line at higher rates indicating blocking loss. (b) System jitter of a 50-detector section. Detection delay was calculated as the time elapsed between the optical pulse being generated and detection event recorded.}
    \label{fig4-timing}
\end{figure}

We additionally measured the timing performance of the array. \reffig{fig4-timing}a shows the count rate from one of the 50-detector sections versus optical attenuation measured at \SI{635}{\nm} and with the detectors biased at 45~\uA per detector. We observed that the photon count rate was proportional to the photon flux, indicating the array was operating in the single-photon regime~\cite{Goltsman2001, PEACOQ}. The count rate was limited primarily by the speed of the bus -- after detection, the current in the bus dropped to zero for a period of time \taudead, and any other photon detections during this time would be lost. For our device this dead time was \SI{9.16}{\us}. Similar to single-pixel SNSPDs, once count-rates approached $1/\taudead$, photons could still be lost due to blocking loss as evidenced by the 3~dB compression point at 1.09\e{5}~cps. To increase the maximum readout rate, additional buses may be added trivially. For a fixed array size, increasing the total number of buses by a factor of $B$ leads to a $B$-fold increase in the maximum readout rate per bus and thus a $B^2$-fold increase in the maximum readout rate of the entire array. 

\reffig{fig4-timing}b shows the timing jitter of one of the array sections. To obtain this data a single section was biased and illuminated with a \SI{1550}{\nm} mode-locked laser with \SI{100}{fs} optical pulse width. The distribution of detection delays had the FWHM of \SI{336}{ps} and standard deviation of \SI{245.4}{ps}. This relatively large jitter can be primarily attributed to the longitudinal geometric jitter of our \SI{5}{mm} long detectors, and could be significantly reduced by converting to a differential-style readout~\cite{Colangelo2023}.

In this work, we demonstrated the largest superconducting photon-counting camera to date by a factor of 20, using an efficient, time-resolved architecture that is both scalable and tolerant of fabrication defects. The architecture allows for the imaging area to contain only detectors, allowing future arrays to reach 100\% fill factors with high system detection effiencies. While the work here demonstrated classical imaging at the single-photon level, we expect future large-format implementations to greatly enable applications in quantum imaging\cite{Moreau2019} such as sub-Rayleigh super-resolution imaging\cite{Giovannetti2009} or sub-shot-noise imaging\cite{Brida2010}.  These applications typically require the detection of quantum correlations in both time and space, and depend heavily on high efficiencies and low background noise.

\section*{Methods}

The chip was fabricated with 6 layers: two WSi detectors layers, two Nb wiring layers, a resistor layer, and a dielectric spacer layer. First, we sputtered a thin (\SI{4}{\nm}) WSi film on a \SI{75}{mm} Si wafer with \SI{150}{\nm} of thermal oxide. The film was in-situ  capped with \SI{2}{\nm} of aSi to avoid oxidation. Then column detectors, series inductors and the readout bus were patterned using SPR660 and optical lithography with an i-line (\SI{365}{\nm}) stepper. Finally these patterns were etched with an inductively-coupled plasma (ICP) etcher using an AR:SF6 etch. Next, a liftoff pattern was created using optical lithography and \SI{90}{\nm} thick Nb wires with \SI{10}{\nm} of PdAu capping were sputtered to create a low inductance bottom Nb wiring layer.

At this point, it is important to planarize the surface before depositing row detectors. In our tests, it was observed that if this planarization step is skipped, the topography of the column detectors results in a step-like edges on the row detectors which reduces the critical current of row detectors, degrading their performance. The planarization can be achieved by using a dilute divinylsiloxane-benzocyclobutene (DVS-BCB) polymer.  This BCB polymer was applied to the wafer, and spun dry at \SI{4000}{rpm}. The wafer was later baked at \SI{250}{C} in a nitrogen furnace for 60~min, to cure the BCB. Later, \SI{25}{\nm} of SiO$_2$ was sputtered on the wafer with \SI{4}{\nm} of amorphous silicon sputtered in-situ as a capping layer. Finally, the wafer was patterned and etched with a CHF$_3$ and O$_2$ chemistry to create vias through the dielectric layer. The PdAu deposited on the bottom Nb wiring layer acts as a non-oxidizing etch stop, allowing us to overetch and create $1\times1$~\micron vias.

Next, \SI{4}{\nm} of WSi film was sputtered again, patterned and etched to produce row detectors and the heaters on the thermal coupler. Later, \SI{30}{\nm} of PdAu was sputtered with a liftoff process to create all of the resistors needed for the device operation (see \reffig{fig2-circuit}). Finally, 200~nm of Nb was  sputtered and patterned with a liftoff process to create the top Nb wiring layer and the GND plane for readout buses. \SI{2}{\nm} of Ti was used as an adhesion layer for the Nb and PdAu layers.

After this we performed the final etch to disconnect the ground from the readout bus, which was intentionally left connected to the ground plane. Skipping this step usually causes antenna-like RF-pickup, leading to damage of the readout bus in our devices. To improve the yield  and reproducibility of lithographically defined patterns, the detectors were designed as images containing 10 detectors, and the stepper was used to repeat this pattern forming the full array.  The readout circuits and readout bus were designed to connect to 50 detectors each. The overall pattern was generated by rotating the wafer inside the stepper and repeating these images, hence both column and row detectors and their readout circuitry are produced with the same reticles.

The chip was fabricated as a $1000\times1000$ array, and during testing we identified a uniform section of 800 column detectors and 500 row detectors, which we designated as the active detection area. This demonstrates the flexibility of the architecture, since the unbiased sections of the array do not waste any bandwidth on the readout buses. The same technique can be straightforwardly applied to allow selection of an active area on-the-fly. 

\subsection*{Circuit parameter values}
In this implementation, the values for the circuit elements were optimized using test structures as well as SPICE simulations. The section biasing resistor \Rbias was 1000~\Ohm, the detector biasing resistor \Rs was 80~\Ohm, the shunting resistor \Rshunt was 16~\Ohm, the thermal coupler series resistor \Rtc was 16~\Ohm, the series inductor \Lseries was \SI{1.25}{\mu H}, and the inductance of the SNSPD \Lsnspd was \SI{1.14}{\mu H}. The biasing resistor \Rs needed to be sufficiently large to ensure that current was distributed uniformly to each detector -- if made too small, parasitic resistances in the wiring or vias can cause uneven distribution.  However, larger values  also contribute to the static power consumption of the array and so its value cannot be too large. $\Rshunt$ was chosen to be relatively small, as it allows the current to be shunted to ground after a detector clicked. By choosing an appropriately small value, the detectors are sufficiently shunted such that latching is impossible, simplifying the operation of the array. Additionally, a small value of $\Rshunt$ guarantees that voltages generated during the detection process are shunted before they reach the communal bias line, minimizing crosstalk between neighboring detectors in a section.

A large value of \Lseries serves to both minimize crosstalk voltages on the bias line during detection, and acts as an energy reservoir with which to power the heating element on the thermal coupler. Too-large values, however, result in wasted space on the chip as well as generate excess heating of the bus that can negatively affect reset times. While smaller $\Lsnspd$ values would be preferred to minimize geometric jitter and reset time in the detectors, for this design we required least \SI{5}{mm}-long detectors to span the imaging area.

\section{Data availability}
Unprocessed time tags and data used to produce images and graphs shown in \reffig{fig1-overview},\reffig{fig3-histograms}, and \reffig{fig4-timing} are available from the corresponding author on reasonable request.

\section{Code availability}
Code used to process the time tags and produce images shown in \reffig{fig1-overview} and \reffig{fig4-timing} is available from the corresponding author on reasonable request.

\section{Acknowledgements}
The U.S. Government is authorized to reproduce and distribute reprints for governmental purposes notwithstanding any copyright annotation thereon. Part of this research was performed at the Jet Propulsion Laboratory, California Institute of Technology, under contract with NASA (Contract No. 80NM0018D0004). A.N.M. was supported in part by NASA APRA via Grant No. NNH17ZDA001N. Support for this work was provided in part by the DARPA DSO Invisible Headlights program. Certain equipment, instruments, software, or materials, commercial or non-commercial, are identified in this paper in order to specify the experimental procedure adequately. Such identification is not intended to imply recommendation or endorsement of any product or service by NIST, nor is it intended to imply that the materials or equipment identified are necessarily the best available for the purpose. This research was funded by NIST (https://ror.org/05xpvk416), University of Colorado Boulder (https://ror.org/02ttsq026), and the Jet Propulsion Laboratory (https://ror.org/027k65916).

\section{Author contributions}
ANM, BK, and BGO conceptualized these experiments. Fabrication of devices was done by BGO. Measurements were performed by BGO, DSR and ANM. Analysis and interpretation of the data was done by BGO, SWN, MDS, BK, and JA. ANM directed and supervised this work. BGO and ANM prepared the manuscript with input from all co-authors.


\newpage

\printbibliography

@article{Riechers2013,
  title    = "A dust-obscured massive maximum-starburst galaxy at a redshift of
              6.34",
  author   = "Riechers, Dominik A and Bradford, C M and Clements, D L and
              Dowell, C D and P{\'e}rez-Fournon, I and Ivison, R J and Bridge,
              C and Conley, A and Fu, Hai and Vieira, J D and Wardlow, J and
              Calanog, J and Cooray, A and Hurley, P and Neri, R and
              Kamenetzky, J and Aguirre, J E and Altieri, B and Arumugam, V and
              Benford, D J and B{\'e}thermin, M and Bock, J and Burgarella, D
              and Cabrera-Lavers, A and Chapman, S C and Cox, P and Dunlop, J S
              and Earle, L and Farrah, D and Ferrero, P and Franceschini, A and
              Gavazzi, R and Glenn, J and Solares, E A Gonzalez and Gurwell, M
              A and Halpern, M and Hatziminaoglou, E and Hyde, A and Ibar, E
              and Kov{\'a}cs, A and Krips, M and Lupu, R E and Maloney, P R and
              Martinez-Navajas, P and Matsuhara, H and Murphy, E J and Naylor,
              B J and Nguyen, H T and Oliver, S J and Omont, A and Page, M J
              and Petitpas, G and Rangwala, N and Roseboom, I G and Scott, D
              and Smith, A J and Staguhn, J G and Streblyanska, A and Thomson,
              A P and Valtchanov, I and Viero, M and Wang, L and Zemcov, M and
              Zmuidzinas, J",
  journal  = "Nature",
  volume   =  496,
  number   =  7445,
  pages    = "329--333",
  month    =  apr,
  year     =  2013,
  language = "en"
}

@article{DeBernardis2000,
  title    = "A flat Universe from high-resolution maps of the cosmic microwave
              background radiation",
  author   = "{de Bernardis P} and Ade, P A and Bock, J J and Bond, J R and
              Borrill, J and Boscaleri, A and Coble, K and Crill, B P and {De
              Gasperis G} and Farese, P C and Ferreira, P G and Ganga, K and
              Giacometti, M and Hivon, E and Hristov, V V and Iacoangeli, A and
              Jaffe, A H and Lange, A E and Martinis, L and Masi, S and Mason,
              P V and Mauskopf, P D and Melchiorri, A and Miglio, L and
              Montroy, T and Netterfield, C B",
  journal  = "Nature",
  volume   =  404,
  number   =  6781,
  pages    = "955--959",
  month    =  apr,
  year     =  2000,
  language = "en"
}

@article{Morozov2021,
author = {Dmitry V. Morozov and Alessandro Casaburi and Robert H. Hadfield},
title = {Superconducting photon detectors},
journal = {Contemporary Physics},
volume = {62},
number = {2},
pages = {69-91},
year  = {2021},
publisher = {Taylor & Francis},
doi = {10.1080/00107514.2022.2043596},
URL = { 
        https://doi.org/10.1080/00107514.2022.2043596
},
eprint = { 
        https://doi.org/10.1080/00107514.2022.2043596
}
}

@article{Wollman2017,
  title   = "{UV} superconducting nanowire single-photon detectors with high
             efficiency, low noise, and 4 {K} operating temperature",
  author  = "Wollman, E E and Verma, V B and Beyer, A D and Briggs, R M and
             Marsili, F and Allmaras, J P and Lita, A E and Mirin, R P and Nam,
             S W and Shaw, M D",
  journal = "Opt. Express",
  volume  =  25,
  number  =  22,
  pages   = "26792",
  year    =  2017
}

@article{Todaro2021,
	title = {State {Readout} of a {Trapped} {Ion} {Qubit} {Using} a {Trap}-{Integrated} {Superconducting} {Photon} {Detector}},
	volume = {126},
	issn = {0031-9007, 1079-7114},
	url = {https://link.aps.org/doi/10.1103/PhysRevLett.126.010501},
	doi = {10.1103/PhysRevLett.126.010501},
	language = {en},
	number = {1},
	urldate = {2023-05-25},
	journal = {Physical Review Letters},
	author = {Todaro, S. L. and Verma, V. B. and McCormick, K. C. and Allcock, D. T. C. and Mirin, R. P. and Wineland, D. J. and Nam, S. W. and Wilson, A. C. and Leibfried, D. and Slichter, D. H.},
	month = jan,
	year = {2021},
	pages = {010501},
}

@article{Arrazola2021,
  title    = "Quantum circuits with many photons on a programmable nanophotonic
              chip",
  author   = "Arrazola, J M and Bergholm, V and Br{\'a}dler, K and Bromley, T R
              and Collins, M J and Dhand, I and Fumagalli, A and Gerrits, T and
              Goussev, A and Helt, L G and Hundal, J and Isacsson, T and
              Israel, R B and Izaac, J and Jahangiri, S and Janik, R and
              Killoran, N and Kumar, S P and Lavoie, J and Lita, A E and
              Mahler, D H and Menotti, M and Morrison, B and Nam, S W and
              Neuhaus, L and Qi, H Y and Quesada, N and Repingon, A and
              Sabapathy, K K and Schuld, M and Su, D and Swinarton, J and
              Sz{\'a}va, A and Tan, K and Tan, P and Vaidya, V D and Vernon, Z
              and Zabaneh, Z and Zhang, Y",
  journal  = "Nature",
  volume   =  591,
  number   =  7848,
  pages    = "54--60",
  month    =  mar,
  year     =  2021,
  language = "en"
}

@article{Verma2021,
  title     = "Single-photon detection in the mid-infrared up to 10 $\mu$m
               wavelength using tungsten silicide superconducting nanowire
               detectors",
  author    = "Verma, V B and Korzh, B and Walter, A B and Lita, A E and
               Briggs, R M and Colangelo, M and Zhai, Y and Wollman, E E and
               Beyer, A D and Allmaras, J P and Vora, H and Zhu, D and Schmidt,
               E and Kozorezov, A G and Berggren, K K and Mirin, R P and Nam, S
               W and Shaw, M D",
  journal   = "APL Photonics",
  publisher = "AIP Publishing, LLC",
  volume    =  6,
  number    =  5,
  pages     = "056101",
  year      =  2021
}

@article{Hadfield2009,
  title     = "Single-photon detectors for optical quantum information
               applications",
  author    = "Hadfield, Robert H",
  journal   = "Nat. Photonics",
  publisher = "Nature Publishing Group",
  volume    =  3,
  number    =  12,
  pages     = "696--705",
  month     =  dec,
  year      =  2009
}

@article{Marsili2013,
  title     = "Detecting single infrared photons with 93 \% system efficiency",
  author    = "Marsili, F and Verma, V B and Stern, J A and Harrington, S and
               Lita, A E and Gerrits, T and Vayshenker, I and Baek, B and Shaw,
               M D and Mirin, R P and Nam, S W",
  publisher = "Nature Publishing Group",
  volume    =  7,
  number    = "February",
  year      =  2013
}

@article{Colangelo2023,
  title = {Impedance-Matched Differential Superconducting Nanowire Detectors},
  author = {Colangelo, Marco and Korzh, Boris and Allmaras, Jason P. and Beyer, Andrew D. and Mueller, Andrew S. and Briggs, Ryan M. and Bumble, Bruce and Runyan, Marcus and Stevens, Martin J. and McCaughan, Adam N. and Zhu, Di and Smith, Stephen and Becker, Wolfgang and Narv\'aez, Lautaro and Bienfang, Joshua C. and Frasca, Simone and Velasco, Angel E. and Ramirez, Edward E. and Walter, Alexander B. and Schmidt, Ekkehart and Wollman, Emma E. and Spiropulu, Maria and Mirin, Richard and Nam, Sae Woo and Berggren, Karl K. and Shaw, Matthew D.},
  journal = {Phys. Rev. Appl.},
  volume = {19},
  issue = {4},
  pages = {044093},
  numpages = {19},
  year = {2023},
  month = {Apr},
  publisher = {American Physical Society},
  doi = {10.1103/PhysRevApplied.19.044093},
  url = {https://link.aps.org/doi/10.1103/PhysRevApplied.19.044093}
}

@ARTICLE{Grunenfelder2023,
  title    = "Fast single-photon detectors and real-time key distillation
              enable high secret-key-rate quantum key distribution systems",
  author   = "Gr{\"u}nenfelder, Fadri and Boaron, Alberto and Resta, Giovanni V
              and Perrenoud, Matthieu and Rusca, Davide and Barreiro, Claudio
              and Houlmann, Rapha{\"e}l and Sax, Rebecka and Stasi, Lorenzo and
              El-Khoury, Sylvain and H{\"a}nggi, Esther and Bosshard, Nico and
              Bussi{\`e}res, F{\'e}lix and Zbinden, Hugo",
  journal  = "Nat. Photonics",
  volume   =  17,
  number   =  5,
  pages    = "422--426",
  month    =  mar,
  year     =  2023,
  language = "en"
}

@ARTICLE{Li2023,
  title     = "High-rate quantum key distribution exceeding 110 Mb s--1",
  author    = "Li, Wei and Zhang, Likang and Tan, Hao and Lu, Yichen and Liao,
               Sheng-Kai and Huang, Jia and Li, Hao and Wang, Zhen and Mao,
               Hao-Kun and Yan, Bingze and Li, Qiong and Liu, Yang and Zhang,
               Qiang and Peng, Cheng-Zhi and You, Lixing and Xu, Feihu and Pan,
               Jian-Wei",
  journal   = "Nat. Photonics",
  publisher = "Nature Publishing Group",
  volume    =  17,
  number    =  5,
  pages     = "416--421",
  month     =  mar,
  year      =  2023,
  language  = "en"
}

@article{Dixit2021,
	title = {Searching for {Dark} {Matter} with a {Superconducting} {Qubit}},
	volume = {126},
	issn = {0031-9007, 1079-7114},
	url = {https://link.aps.org/doi/10.1103/PhysRevLett.126.141302},
	doi = {10.1103/PhysRevLett.126.141302},
	language = {en},
	number = {14},
	urldate = {2023-05-25},
	journal = {Physical Review Letters},
	author = {Dixit, Akash V. and Chakram, Srivatsan and He, Kevin and Agrawal, Ankur and Naik, Ravi K. and Schuster, David I. and Chou, Aaron},
	month = apr,
	year = {2021},
	pages = {141302},
}

@ARTICLE{Wang2022,
  title     = "Twin-field quantum key distribution over 830-km fibre",
  author    = "Wang, Shuang and Yin, Zhen-Qiang and He, De-Yong and Chen, Wei
               and Wang, Rui-Qiang and Ye, Peng and Zhou, Yao and Fan-Yuan,
               Guan-Jie and Wang, Fang-Xiang and Zhu, Yong-Gang and Morozov,
               Pavel V and Divochiy, Alexander V and Zhou, Zheng and Guo,
               Guang-Can and Han, Zheng-Fu",
  journal   = "Nat. Photonics",
  publisher = "Nature Publishing Group",
  volume    =  16,
  number    =  2,
  pages     = "154--161",
  month     =  jan,
  year      =  2022,
  language  = "en"
}

@ARTICLE{Boaron2018,
  title   = "Secure Quantum Key Distribution over 421 km of Optical Fiber",
  author  = "Boaron, Alberto and Boso, Gianluca and Rusca, Davide and Autebert,
             Claire and Caloz, Misael and Perrenoud, Matthieu and Gras,
             Ga{\"e}tan and Li, Ming-Jun and Nolan, Daniel and Martin, Anthony
             and Zbinden, Hugo",
  journal = "Phys. Rev. Lett.",
  volume  =  121,
  pages   = "190502",
  year    =  2018
}

@ARTICLE{Wollman2019,
  title    = "A kilopixel array of superconducting nanowire single-photon
              detectors",
  author   = "Wollman, Emma E and Verma, Varun B and Lita, Adriana E and Farr,
              William H and Shaw, Matthew D and Mirin, Richard P and Nam, Sae
              Woo",
  journal  = "Opt. Express",
  volume   =  27,
  number   =  24,
  pages    = "35279--35289",
  year     =  2019
}

@article{PEACOQ,
author = {Ioana Craiciu and Boris Korzh and Andrew D. Beyer and Andrew Mueller and Jason P. Allmaras and Lautaro Narv\'{a}ez and Maria Spiropulu and Bruce Bumble and Thomas Lehner and Emma E. Wollman and Matthew D. Shaw},
journal = {Optica},
number = {2},
pages = {183--190},
publisher = {Optica Publishing Group},
title = {High-speed detection of 1550 nm single photons with superconducting nanowire detectors},
volume = {10},
month = {Feb},
year = {2023},
url = {https://opg.optica.org/optica/abstract.cfm?URI=optica-10-2-183},
doi = {10.1364/OPTICA.478960},
}

@article{Goltsman2001,
    author = {Gol’tsman, G. N. and Okunev, O. and Chulkova, G. and Lipatov, A. and Semenov, A. and Smirnov, K. and Voronov, B. and Dzardanov, A. and Williams, C. and Sobolewski, Roman},
    title = "{Picosecond superconducting single-photon optical detector}",
    journal = {Applied Physics Letters},
    volume = {79},
    number = {6},
    pages = {705-707},
    year = {2001},
    month = {08},
    issn = {0003-6951},
    doi = {10.1063/1.1388868},
    url = {https://doi.org/10.1063/1.1388868},
    eprint = {https://pubs.aip.org/aip/apl/article-pdf/79/6/705/8782545/705\_1\_online.pdf},
}

@article{dcrexpcurve,
    author = {Yamashita, T. and Miki, S. and Makise, K. and Qiu, W. and Terai, H. and Fujiwara, M. and Sasaki, M. and Wang, Z.},
    title = "{Origin of intrinsic dark count in superconducting nanowire single-photon detectors}",
    journal = {Applied Physics Letters},
    volume = {99},
    number = {16},
    year = {2011},
    month = {10},
    issn = {0003-6951},
    doi = {10.1063/1.3652908},
    url = {https://doi.org/10.1063/1.3652908},
    note = {161105},
    eprint = {https://pubs.aip.org/aip/apl/article-pdf/doi/10.1063/1.3652908/14458078/161105\_1\_online.pdf},
}

@article{ Nones2012,
	title = {High-impedance {NbSi} {TES} sensors for studying the cosmic microwave background radiation},
	volume = {548},
	issn = {0004-6361, 1432-0746},
	url = {http://www.aanda.org/10.1051/0004-6361/201218834},
	doi = {10.1051/0004-6361/201218834},
	urldate = {2023-05-25},
	journal = {Astronomy \& Astrophysics},
	author = {Nones, C. and Marnieros, S. and Benoit, A. and Bergé, L. and Bideaud, A. and Camus, P. and Dumoulin, L. and Monfardini, A. and Rigaut, O.},
	month = dec,
	year = {2012},
	pages = {A17},
}

@article{Dileep98,
author = {Dileep V. Reddy and Robert R. Nerem and Sae Woo Nam and Richard P. Mirin and Varun B. Verma},
journal = {Optica},
number = {12},
pages = {1649--1653},
publisher = {Optica Publishing Group},
title = {Superconducting nanowire single-photon detectors with 98$\%$ system detection efficiency at 1550 nm},
volume = {7},
month = {Dec},
year = {2020},
url = {https://opg.optica.org/optica/abstract.cfm?URI=optica-7-12-1649},
doi = {10.1364/OPTICA.400751},
}

@article{Boris3ps,
author = {Boris Korzh and Qing-Yuan Zhao and Jason P. Allmaras and Simone Frasca and Travis M. Autry and Eric A. Bersin and Andrew D. Beyer and Ryan M. Briggs and Bruce Bumble and Marco Colangelo and Garrison M. Crouch and Andrew E. Dane and Thomas Gerrits and Adriana E. Lita and Francesco Marsili and Galan Moody and Cristian Pena and Edward Ramirez and Jake D. Rezac and Neil Sinclair and Martin J. Stevens and Angel E. Velasco and Varun B. Verma and Emma E. Wollman and Si Xie and Di Zhu and Paul D. Hale and Maria Spiropulu, Kevin L. Silverman and Richard P. Mirin and Sae Woo Nam, Alexander G. Kozorezov and Matthew D. Shaw and Karl K. Berggren},
journal = {Nature Photonics},
volume = {14},
issue = {4},
pages = {250–255},
title = {Demonstration of sub-3 ps temporal resolution with a superconducting nanowire single-photon detector},
year = {2020},
url = {https://doi.org/10.1038/s41566-020-0589-x},
doi = {https://doi.org/10.1038/s41566-020-0589-x},
}

@article{VarunRowCol,
author = {Verma,V. B.  and Horansky,R.  and Marsili,F.  and Stern,J. A.  and Shaw,M. D.  and Lita,A. E.  and Mirin,R. P.  and Nam,S. W. },
title = {A four-pixel single-photon pulse-position array fabricated from WSi superconducting nanowire single-photon detectors},
journal = {Applied Physics Letters},
volume = {104},
number = {5},
pages = {051115},
year = {2014},
doi = {10.1063/1.4864075},
URL = {https://doi.org/10.1063/1.4864075},
eprint = {https://doi.org/10.1063/1.4864075}}

@article{Zhao2017,
author = {Qing-Yuan Zhao and Di Zhu and Niccolo Calandri and Andrew E. Dane and Adam N. McCaughan and Francesco Bellei and Hao-Zhu Wang and Daniel F. Santavicca and Karl K. Berggren},
journal = {Nature Photonics},
volume = {11},
issue = {4},
pages = {247–251},
title = {Single-photon imager based on a superconducting nanowire delay line},
year = {2017},
url = {https://doi.org/10.1038/nphoton.2017.35},
doi = {https://doi.org/10.1038/nphoton.2017.35},
}

@article{chilesdcr,
  title = {New Constraints on Dark Photon Dark Matter with Superconducting Nanowire Detectors in an Optical Haloscope},
  author = {Chiles, Jeff and Charaev, Ilya and Lasenby, Robert and Baryakhtar, Masha and Huang, Junwu and Roshko, Alexana and Burton, George and Colangelo, Marco and Van Tilburg, Ken and Arvanitaki, Asimina and Nam, Sae Woo and Berggren, Karl K.},
  journal = {Phys. Rev. Lett.},
  volume = {128},
  issue = {23},
  pages = {231802},
  numpages = {7},
  year = {2022},
  month = {Jun},
  publisher = {American Physical Society},
  doi = {10.1103/PhysRevLett.128.231802},
  url = {https://link.aps.org/doi/10.1103/PhysRevLett.128.231802}
}

@article{Brida2010,
	title = {Experimental realization of sub-shot-noise quantum imaging},
	volume = {4},
	issn = {1749-4885, 1749-4893},
	url = {https://www.nature.com/articles/nphoton.2010.29},
	doi = {10.1038/nphoton.2010.29},
	language = {en},
	number = {4},
	urldate = {2023-05-25},
	journal = {Nature Photonics},
	author = {Brida, G. and Genovese, M. and Ruo Berchera, I.},
	month = apr,
	year = {2010},
	pages = {227--230},
}

@article{Moreau2019,
	title = {Imaging with quantum states of light},
	volume = {1},
	issn = {2522-5820},
	url = {https://www.nature.com/articles/s42254-019-0056-0},
	doi = {10.1038/s42254-019-0056-0},
	language = {en},
	number = {6},
	urldate = {2023-05-25},
	journal = {Nature Reviews Physics},
	author = {Moreau, Paul-Antoine and Toninelli, Ermes and Gregory, Thomas and Padgett, Miles J.},
	month = may,
	year = {2019},
	pages = {367--380},
}

@article{Giovannetti2009,
	title = {Sub-{Rayleigh}-diffraction-bound quantum imaging},
	volume = {79},
	issn = {1050-2947, 1094-1622},
	url = {https://link.aps.org/doi/10.1103/PhysRevA.79.013827},
	doi = {10.1103/PhysRevA.79.013827},
	language = {en},
	number = {1},
	urldate = {2023-05-25},
	journal = {Physical Review A},
	author = {Giovannetti, Vittorio and Lloyd, Seth and Maccone, Lorenzo and Shapiro, Jeffrey H.},
	month = jan,
	year = {2009},
	pages = {013827},
}

@article{Madsen2022,
	title = {Quantum computational advantage with a programmable photonic processor},
	volume = {606},
	issn = {0028-0836, 1476-4687},
	url = {https://www.nature.com/articles/s41586-022-04725-x},
	doi = {10.1038/s41586-022-04725-x},
	language = {en},
	number = {7912},
	urldate = {2023-05-25},
	journal = {Nature},
	author = {Madsen, Lars S. and Laudenbach, Fabian and Askarani, Mohsen Falamarzi. and Rortais, Fabien and Vincent, Trevor and Bulmer, Jacob F. F. and Miatto, Filippo M. and Neuhaus, Leonhard and Helt, Lukas G. and Collins, Matthew J. and Lita, Adriana E. and Gerrits, Thomas and Nam, Sae Woo and Vaidya, Varun D. and Menotti, Matteo and Dhand, Ish and Vernon, Zachary and Quesada, Nicolás and Lavoie, Jonathan},
	month = jun,
	year = {2022},
	pages = {75--81},
}

@article{varun1kpx,
author = {Emma E. Wollman and Varun B. Verma and Adriana E. Lita and William H. Farr and Matthew D. Shaw and Richard P. Mirin and Sae Woo Nam},
journal = {Opt. Express},
number = {24},
pages = {35279--35289},
publisher = {Optica Publishing Group},
title = {Kilopixel array of superconducting nanowire single-photon detectors},
volume = {27},
month = {Nov},
year = {2019},
url = {https://opg.optica.org/oe/abstract.cfm?URI=oe-27-24-35279},
doi = {10.1364/OE.27.035279},
}

@article{Jasonthermal,
author = {Allmaras, Jason P. and Wollman, Emma E. and Beyer, Andrew D. and Briggs, Ryan M. and Korzh, Boris A. and Bumble, Bruce and Shaw, Matthew D.},
title = {Demonstration of a Thermally Coupled Row-Column SNSPD Imaging Array},
journal = {Nano Letters},
volume = {20},
number = {3},
pages = {2163-2168},
year = {2020},
doi = {10.1021/acs.nanolett.0c00246},
    note ={PMID: 32091221},
URL = {https://doi.org/10.1021/acs.nanolett.0c00246 },
eprint = { https://doi.org/10.1021/acs.nanolett.0c00246   
}
}

@article{tci1kpx,
author = {McCaughan,A. N.  and Zhai,Y.  and Korzh,B.  and Allmaras,J. P.  and Oripov,B. G.  and Shaw,M. D.  and Nam,S. W. },
title = {The thermally coupled imager: A scalable readout architecture for superconducting nanowire single photon detectors},
journal = {Applied Physics Letters},
volume = {121},
number = {10},
pages = {102602},
year = {2022},
doi = {10.1063/5.0102154},
URL = {https://doi.org/10.1063/5.0102154},
eprint = {https://doi.org/10.1063/5.0102154}
}

@article{Adamhtron,
author = {McCaughan, A. N. and Verma, V. B. and Buckley, S. M. and Allmaras, J. P. and Kozorezov, A. G. and Tait, A. N. and Nam, S. W. and Shainline, J. M.},
title = {A superconducting thermal switch with ultrahigh impedance for interfacing superconductors to semiconductors},
journal = {Nature Electronics},
volume = {2},
number = {10},
pages = {451–456},
year = {2019},
doi = {10.1038/s41928-019-0300-8},
URL = {https://doi.org/10.1038/s41928-019-0300-8},
eprint = {https://doi.org/10.1038/s41928-019-0300-8}
}

@article{belltest,
  title = {Strong Loophole-Free Test of Local Realism},
  author = {Shalm, Lynden K. and Meyer-Scott, Evan and Christensen, Bradley G. and Bierhorst, Peter and Wayne, Michael A. and Stevens, Martin J. and Gerrits, Thomas and Glancy, Scott and Hamel, Deny R. and Allman, Michael S. and Coakley, Kevin J. and Dyer, Shellee D. and Hodge, Carson and Lita, Adriana E. and Verma, Varun B. and Lambrocco, Camilla and Tortorici, Edward and Migdall, Alan L. and Zhang, Yanbao and Kumor, Daniel R. and Farr, William H. and Marsili, Francesco and Shaw, Matthew D. and Stern, Jeffrey A. and Abell\'an, Carlos and Amaya, Waldimar and Pruneri, Valerio and Jennewein, Thomas and Mitchell, Morgan W. and Kwiat, Paul G. and Bienfang, Joshua C. and Mirin, Richard P. and Knill, Emanuel and Nam, Sae Woo},
  journal = {Phys. Rev. Lett.},
  volume = {115},
  issue = {25},
  pages = {250402},
  numpages = {10},
  year = {2015},
  month = {Dec},
  publisher = {American Physical Society},
  doi = {10.1103/PhysRevLett.115.250402},
  url = {https://link.aps.org/doi/10.1103/PhysRevLett.115.250402}
}

@article{Giustina2015,
author = {Giustina, Marissa and Versteegh, Marijn A. M. and Wengerowsky, S{\"{o}}ren and Handsteiner, Johannes and Hochrainer, Armin and Phelan, Kevin and Steinlechner, Fabian and Kofler, Johannes and Larsson, Jan-{\AA}ke and Abell{\'{a}}n, Carlos and Amaya, Waldimar and Pruneri, Valerio and Mitchell, Morgan W. and Beyer, J{\"{o}}rn and Gerrits, Thomas and Lita, Adriana E. and Shalm, Lynden K. and Nam, Sae Woo and Scheidl, Thomas and Ursin, Rupert and Wittmann, Bernhard and Zeilinger, Anton},
doi = {10.1103/PhysRevLett.115.250401},
issn = {0031-9007},
journal = {Physical Review Letters},
month = {dec},
number = {25},
pages = {250401},
title = {{Significant-Loophole-Free Test of Bell's Theorem with Entangled Photons}},
url = {https://link.aps.org/doi/10.1103/PhysRevLett.115.250401},
volume = {115},
year = {2015}
}

@article{Walter2020,
author = {Walter, Alexander B. and Fruitwala, Neelay and Steiger, Sarah and Bailey, John I. and Zobrist, Nicholas and Swimmer, Noah and Lipartito, Isabel and Smith, Jennifer Pearl and Meeker, Seth R. and Bockstiegel, Clint and Coiffard, Gregoire and Dodkins, Rupert and Szypryt, Paul and Davis, Kristina K. and Daal, Miguel and Bumble, Bruce and Collura, Giulia and Guyon, Olivier and Lozi, Julien and Vievard, Sebastien and Jovanovic, Nemanja and Martinache, Frantz and Currie, Thayne and Mazin, Benjamin A.},
doi = {10.1088/1538-3873/abc60f},
issn = {1538-3873},
journal = {Publications of the Astronomical Society of the Pacific},
month = {nov},
number = {1018},
pages = {125005},
title = {{The MKID Exoplanet Camera for Subaru SCExAO}},
url = {https://iopscience.iop.org/article/10.1088/1538-3873/abc60f},
volume = {132},
year = {2020}
}

@article{Day2003a,
author = {Day, Peter K. P.K. and LeDuc, Henry G. H.G. and Mazin, Benjamin A. and Vayonakis, Anastasios and Zmuidzinas, Jonas},
doi = {10.1038/nature02037},
issn = {0028-0836},
journal = {Nature},
month = {oct},
number = {6960},
pages = {817--821},
publisher = {Nature Publishing Group},
title = {{A broadband superconducting detector suitable for use in large arrays}},
url = {http://www.nature.com/nature/journal/v425/n6960/abs/nature02037.html http://dx.doi.org/10.1038/nature02037 http://www.nature.com/nature/journal/v425/n6960/full/nature02037.html},
volume = {425},
year = {2003}
}

\end{document}